\documentclass{svjour3}
\usepackage{graphicx}
\usepackage{latexsym}
\usepackage{caption}
\usepackage{amssymb}
\smartqed
\newcommand{\ti}[1]{\mbox{\tiny{#1}}}

\newcommand{\ap}[1]{\ ^{\mbox{\tiny{(#1)}}}\! }
\begin{document}
\title
{Massive test particles motion in Kaluza-Klein gravity}
\author{ Valentino Lacquaniti \and Giovanni Montani\and Daniela Pugliese}
\institute{ Valentino Lacquaniti \at
 Physics Department ``E.Amaldi'', University of Rome, "'Roma Tre"',
Via della Vasca Navale 84, I-00146, Rome , Italy\\
\email{lacquaniti@fis.uniroma3.it \and Giovanni Montani \at ICRA
International Center for Relativistic Astrophysics, Physics
Department (G9), University of Rome, ``La Sapienza'',
Piazzale Aldo Moro 5, 00185 Rome, Italy   \\
 ENEA-C. R. Frascati Unita' Fus. Mag., via E.Fermi 45,
I-00044, Frascati, Rome, Italy \\
\email{Montani@icra.it} \and Daniela Pugliese \at ICRA International
Center for Relativistic Astrophysics, Physics Department (G9),
University of Rome, ``La Sapienza'', Piazzale Aldo Moro 5, 00185
Rome, Italy
\\
\email{Pugliese@icra.it}}}
\date{Received: date / Accepted: date}
\maketitle
\begin{abstract}
A class of static, vacuum  solutions of (free-electromagnetic)
Kaluza-Klein equations with three-dimensional spherical symmetry is
studied.  In order to explore the dynamic in such spacetimes,
geodesic equations are obtained and the effective potential for
massive test particles is analysed. Particular attention is devoted
to the properties of the four-dimensional counterpart of these
solutions in their Schwarzschild limit. A modification of the
circular stable orbits compared with the Schwarzschild case is
investigated. 

\end{abstract}

\maketitle
\section{Introduction}
 Extra dimensional theories are
candidate for the great unification, being  based on the effort to
extend to other fields the geometrical picture of gravitation
(\cite{Bergamini:1984gx}, \cite{Aranda:2009wh}). Indeed, some
cosmological models,  including for instance  strings or brane
worlds, take an implement of the number of dimensions of the
spacetime  to the five of the original Kaluza Klein model or more
dimensions (\cite{Montani:2009wg}, \cite{Gunther:2002mm},
\cite{Brax:2003fv}, \cite{Langlois:2002bb},
\cite{Papantonopoulos:2002ew}). On one side great interest is
involved to provide a theoretical model able to explain the role
 extra dimensions and their compatibility in a world that looks like a
four dimensional one. On the other side any experimental observation
that could be compatible with such theories could be a strong
constraint concerning their validity. Moreover, scalar fields which
are naturally provided by such theories play  a crucial role in the
dynamics  of the present inflationary models (see for example
\cite{Kolb}). 5D Kaluza Klein models provide the geometrization of
the electromagnetism and a scalar field associated to
 extra dimensional component of the metric.  The gauge invariance
arises as a spacetime symmetry  realized imposing the invariance for
translations on the compactified fifth dimension. Nevertheless, the
study of test particle dynamics shows a great problem of such
theory, known as the charge-mass puzzle. It is possible to recover
the Lorentz equation for the  particle motion, but the charge-mass
ratio does not match with any observed particle because the theory
provide huge massive modes near the Planck scale. Some works propose
a solution making a  revision of the  approach to the particle
dynamics which is usually adopted in these models; for instance in
\cite{Lacquaniti:2009yy}, a definition of a 5D particle as a
localized matter distribution in the ordinary 4D spacetime but as a
delocalized one  on the fifth dimension is considered. It leads to a
different definition of mass that solves the charge-mass puzzle.

 In this work we study test particles motion
in a  five dimensional, electromagnetic-free, Kaluza-Klein (KK)
model. As an extension of the Schwarzschild solution in a 5D
scenario, we consider here a vacuum solution of KK equations with
3D-spherical symmetry. Using an effective potential approach to the
motion, we are able to find the last circular orbit radius and in
particular the last stable circular orbits radius of a charged or
neutral test particle. The work is motivated by the aim to provide
an experimental constraints on the validity of multidimensional
gravity theory exploring  the dynamical effects of the extra
compactified dimension. The presence of such a dimension should
produce a non trivial departure from the dynamics in the
corresponding 4D counterparts of these solutions. At first, we
analyze the test particles motion by a standard approach to the
Kaluza-Klein dynamic, therefore performing a dimensional reduction
to four dimensions of a 5D free particle following a 5D  geodesic.
Then  we compare this approach with the new one realized in
\cite{Lacquaniti:2009yy}, based on a Papapetrou multipole expansion
of a 5D energy-momentum tensor which is supposed to be picked along
a 4D-world tube.

 The paper
is organized as follows: in Sec. \ref{sec:5dgenerico} we review some
fundamental statements of KK model. In
Sec. \ref{sec:particledinamic} we examine test particles motion
reviewing  first the geodesic approach and then considering the
dynamics from
 point of view of the multipole expansion. In Sec. \ref{sec:elogia} we
review the circular  motion of a test particle in Schwarzschild
geometry by mean of the effective potential approach. In Sec.
\ref{sec:GSS} we recall some of the properties of the generalized
Schwarzschild solution. Finally, in Sec.\ref{sec:CircularorGSS} we
explore the dynamics in such spacetimes, we find an effective
potential and we study circular orbits, either in the standard
scheme of   the motion in KK gravity, either in the approach \emph{a
l$\acute{a}$}  Papapetrou. The paper will end in
Sec.\ref{sec:conclusion} where concluding remarks follow.
\section{Five dimensional Kaluza-Klein Model}\label{sec:5dgenerico}
The 5D compactified KK model  is settled by the following
assumptions (see for instance
\cite{Overduin:1998pn},\cite{Bailin:1987jd},\cite{Librogra}). The
5D-manifold $\mathcal{M}^{\ti{5}}$ is a direct product
$\mathcal{M}^{\ti{4}}\otimes \mathcal{S}^{\ti{1}}$, between the
ordinary 4D-spacetime $\mathcal{M}^{\ti{4}}$  and  the space-like
loop  $\mathcal{S}^{\ti{1}}$. To make the extra dimension
unobservable its size is assumed to be
below the present observational bound\footnote{This means %
$$
L_{\ti{(5)}}\equiv\int d^{5}x \sqrt{g_{55}}<10^{-18}cm.
$$
} (Compactification hypothesis). Metrics components  do not depend
on the fifth coordinate (Cylindricity hypothesis): such a scenario
could be realised assuming we are working at the lowest order of the
Fourier expansion along the fifth dimension, providing then an
effective theory. Finally, we assume that the $(55)$- component of
the metrics is a scalar. Such a setup results in a breaking of the
5D covariance and the 5D Equivalence Principle
(\cite{Lacquaniti:2009yy}); noticeably, only traslations along the
fifth dimension are allowed and by this way the abelian gauge
invariance of the electromagnetism is realised in  KK model as a
coordinate transformation in $S^1$. According to the KK reduction
the 5D line element reads\footnote{With latin capital letters $A$ we
label the five-dimensional indices, where they run in
$\{0,1,2,3,5\}$, Greek and latin indices $a$  run from 0 to 3, the
spatial indexes $(i,j)$ in $\{1,2,3\}$. We consider metric of
$\{+,-,-,-,-\}$ signature.}
 as follows:
\begin{equation}\label{cic}
ds^2_{\ti{(5)}}=g_{\mu\nu}dx^{\mu}dx^{\nu}-
\phi^2\left(dx_{\ti{5}}+ekA_{\mu}dx^{\mu}\right)^2\, . \label{zippo}
\end{equation}
We adopt coordinates $x^{\mu}$ for ordinary 4D-spacetime while
$x^{5}$ is the angle parameter for the fifth circular dimension. The
extra scalar field $\phi$ we have in the model is the scale factor
governing the expansion of the extra dimension, being
$\phi^2=-g_{55}$;   $A_{\mu}$ represents the electromagnetic field
and $g_{\mu\nu}$ is the usual 4D metric tensor; $ek$ is a
dimensional constant such that $e^2k^2=(4G)/c^2$. In this work we
just concern our analysis to those electromagnetic free -solutions
$(A_{\mu}=0)$, i.e.we deal with a pure scalar tensor theory (for a
discussion of the role of the scalar field in the KK paradigm see
for example \cite{Damour:1992kf}-\cite{Damour:1996ke}).
\section{Particle dynamics in Kaluza Klein
models}\label{sec:particledinamic}
Here we briefly review the
geodesic approach to motion in KK model and  then  we  discuss the  main
features of  the Papapetrou revised approach to the motion recently
appeared in literature.
\subsection{Geodesic approach to motion in KK model}
Borrowing the formulation of motion from the 4D theory, a first approach is simply to assume that the particle motion is governed by the Action
\begin{equation}\label{Stracciatella1}
S_{\ti{(5)}}\equiv-\mu_{\ti{(5)}}\int ds_{\ti{(5)}},
\end{equation}
where the  mass parameter $\mu_{\ti{(5)}}$ is assumed to be
constant, according to the  assumption of equivalence between the
motion of the particle and the  5D geodesic trajectory (see for
example \cite{Librogra},\cite{Overduin:1998pn}). From  Action
(\ref{Stracciatella1}) the 5D  equations is obtained:
\begin{equation}\label{Stracciatella}
\omega^{\ti{A}}\; {^{\ti{(5)}}}\nabla_{\ti{A}}\omega^{\ti{B}}=0\, .
\end{equation}
Here $^{\ti{(5)}}\!\nabla$ is the covariant derivative compatible
with the 5D-metric. 5D-velocities $\omega^{\ti{A}}$ and 4D
velocities $u^{\ti{A}}$ are defined respectively as $
\omega^{\ti{A}}\equiv dx^{\ti{A}}/ds_{\ti{(5)}}$, $ u^{\ti{A}}\equiv
dx^{\ti{A}}/ds$, with $\omega^{\ti{A}}= \alpha u^{\ti{A}}$ and $
g_{\ti{A}\ti{B}}\omega^{\ti{A}}\omega^{\ti{B}}=1$, $
 g_{ab}u^{a}u^{b}=1$,
 where the  $\alpha$ parameter reads:
$$
\alpha\equiv\frac{ds}{ds_{\ti{(5)}}}=\sqrt{g_{ab}\omega^{a}\omega^{b}}
=\sqrt{1+\frac{\omega_{5}^{2}}{\phi^{2}}}
$$
The dimensional reduction of Eq.\ref{Stracciatella}
  (see also \cite{Lacquaniti:2009yy})
provides the set
\begin{eqnarray}
\label{Revisione} u^{a}\  ^{\ti{(4)}}\!\nabla_{a}u^{b}&=& e
k\left(\frac{\omega_{5}}
{\sqrt{1+\frac{\omega_{5}^{2}}{\phi^{2}}}}\right)
F^{bc}u_{c}+\frac{1}{\phi^{3}}\left(u^{b}u^{c}-g^{bc}\right)\partial_{b}\phi\left(\frac{\omega_{5}}
{\sqrt{1+\frac{\omega_{5}^{2}}{\phi^{2}}}}\right)^{2} \\
\label{Pensione} \frac{d\omega_{5}}{d s} &=& 0
\end{eqnarray}
where $F_{ab}=\partial_{a}A_{b}-\partial_{b}A_{a}$ is the Faraday
tensor. Hence a free 5D-test particle   becomes a 4D-interacting
particle, whose motion is described by (\ref{Revisione}).
Eq.\ref{Pensione} provides a constant of motion in agreement with
the existence of the  Killing vector $(0,0,0,0,1)$. Coupling factors
are indeed
 functions of $\omega_{5}$. In particular the
electrodynamics coupling factors, in terms of the effective particle
 charge-mass ratio $q/\mu_{\ti{(5)}}$ is
\begin{equation}\label{volonta}
\frac{q}{\mu_{\ti{(5)}}}=ek \frac{\omega_{5}}
{\sqrt{1+\frac{\omega_{5}^{2}}{\phi^{2}}}}
\end{equation}
 The right member of the (\ref{volonta}) is in general no constant
and always upper bounded. Particularly, if we set $\phi=1$, in order
to restore the Einstein-Maxwell theory, we have the bound
$q<\mu_{\ti{(5)}}$ which is unacceptable for every known elementary
particle. It could be envisaged how such a problem is related within
the background of the geodesic approach  to the problem of the huge
massive mode of the KK tower
(\cite{Lacquaniti:2009yy,Lacquaniti:2009cr,Lacquaniti:2009rq}).  For
$\omega_{5}=0$, neutral particle test case, Eq. \ref{volonta}
becomes a geodetic one. Moreover, even  in a free-electromagnetic
scenario, as prospected in the GSS case, charged $(\omega_{5}\neq0)$
particles, being coupled with the extra-dimensional scalar field by
a $\omega_{5}$-function do not follow in general a geodetic motion.
\subsection{Papapetrou approach to motion in KK model}
In the geodesic approach to the dynamics, the point-like size of the
test particle in   $\mathcal{M}^{\ti{5}}$ is assumed. This
assumption has been recently discussed in some works
\cite{Lacquaniti:2009yy,Lacquaniti:2009cr,Lacquaniti:2009rq}, where
the validity of a model with a point-like particle in a compactified
dimension is criticised. A new proposal is given, where, adopting a
Papapetrou multipole expansion \cite{papapetrou}, the particle is
described as a localized source in $M^{\ti{4}}$ but still
delocalized along the fifth dimension as a consequence of the
compactification. Introducing a generic energy-momentum tensor
$^{\ti{(5)}}\!\mathcal{T}^{\ti{AB}}$ associated to the body,
governed by conservation laws and not depending on the fifth
coordinate,  like it happens for metric fields, the
 following  equations are considered:
\begin{equation}\label{Amor}
^{\ti{(5)}}\!\nabla_{\ti{A}} \
^{\ti{(5)}}\!\mathcal{T}^{\ti{AB}}=0\quad\partial_{5} \
^{\ti{(5)}}\!\mathcal{T}^{\ti{AB}}=0
\label{wwww}
\end{equation}
Performing a multipole expansion \cite{papapetrou} centrad on a
trajectory $X^a$, at the lowest order the procedure gives the motion
equation for a test particle:
\begin{equation}\label{padredimaria}
m u^{a}\ ^{\ti{(4)}}\!\nabla_{a}u^{b}=
(u^{b}u^{c}-g^{bc})\left(\frac{\partial_{c}\phi}{\phi^{3}}\right)A+q
F^{bc}u_{c}
\end{equation}
Below the definitions for
coupling factor $m$, $q$, $A$ and the according definitions for the effective test-particle tensor component follow:
\begin{eqnarray}
m &=& \frac{1}{u^{0}}\int d^{3}x \sqrt{g} \phi  \
^{\ti{(5)}}\!\mathcal{T}^{00},\quad \phi \sqrt{g} T^{\mu\nu}=\int ds
m \delta^{4}\left(x-X\right) u^{\mu} u^{\nu}
\\\label{qdef}
q &=& e k \int d^{3}x \sqrt{g} \phi  \
^{\ti{(5)}}\!\mathcal{T}_{5}^{0},\quad e k\phi \sqrt{g}
T^{\mu}_{5}=\int ds q\delta^{4}\left(x-X\right) u^{\mu}= \sqrt{g}
J^{\mu}
\\\label{Adef}
A &=& u^{0} \int d^{3}x \sqrt{g} \phi
^{\ti{(5)}}\!\mathcal{T}_{55},\quad\phi \sqrt{g} T_{55}=\int ds A
\delta^{4}\left(x-X\right)
\end{eqnarray}
The parameter $m$ correctly represents the mass of the particle,
which turns out to be localized just in the ordinary 4D space, as it
is envisaged by the presence of a 4D Dirac delta function in the
above definitions. The  equation  \ref{padredimaria} admits an
effective Action which does not coincides to the Action
\ref{Stracciatella1}. Via an Hamiltonian analysis of such a revised
Action, it can be proved that the KK tower of massive modes is
suppressed, and the $q/m$ ratio is no more upper bounded. Indeed, it
can be proved that the motion of the particle is correctly governed
by a dispersion relation of the  form
\begin{equation}
P_{\mu}P^{\mu}=m^2\,,
\label{zanzibar}
\end{equation}
where mass is now variable due to scalar fields.
Therefore such an approach allows to deal with test particle consistently without giving up with the compactification hypothesis. Charge $q$ is still conserved, in consequence of the  continuity equation
 $\nabla_{\mu}J^{\mu}=0$, which arises from (\ref{wwww}).
 Mass is in general not conserved and its behaviour is given by
 \begin{equation} \frac{\partial m}{\partial
x^{\mu}}=-\frac{A}{\phi^{3}}\frac{\partial\phi}{\partial x^{\mu}}\, .
\label{padredisofia}
\end{equation}
Therefore the behaviour of mass is related to the variation of the
scalar field and the new coupling $A$ (which has a pure
extra-dimensional origin) along the path. An interesting scenario
concerning $A$ to be investigated should be to assume $ A\,\infty
\,m\phi^2$: by this way Eq.\ref{padredisofia} admits an easy
integration, providing a power law dependence of mass on the scalar
field and , more important, restoring the free falling universality
of particle in Eq.\ref{padredimaria} when a vanishing
electromagnetic field is considered.
\section{Circular orbits  in a Schwarzschild
space-time}\label{sec:elogia}

Let us consider the usual 4D
Schwarzschild geometry:
\begin{equation}\label{117jse8}
ds^2=\Delta(r)dt^2-\Delta(r)^{-1}dr^2
-r^2\left(d\theta^2+\sin^2\theta d\varphi^2\right)
\end{equation}
where $\Delta(r)=\left(1-2M/r\right)$. Using  background symmetries,
we now  restrict   to the equatorial geodesics. Tangent vector
$u^{\alpha}$ to such a curve is $
 u^{\alpha}=dx^{\alpha}/d\tau=\dot{x}^{\mu}$,
 where we choose $\tau$ to be the proper time.
The metric (\ref{117jse8}) admits the  Killing field
$\xi_{t}=\partial_{t} $ and $\xi_{\varphi}=\partial_{\varphi} $ ,
therefore we have  the constants of the motion $ E
=g_{\alpha\beta}\xi_{t}^{\alpha} u^{\beta} =
\Delta(r)(\dot{t})\quad$ and $L =-
g_{\alpha\beta}\xi_{\varphi}^{\alpha}u^{\beta} =
\dot{r}^2\dot{\varphi} $ that  respectively represent  the total
energy (per unit rest mass) of a particle with respect to a static
observer at infinity, and the angular momentum (per unit rest mass)
of the particle. Given a particle with rest mass $\mu$ its
dispersion relation $g_{\alpha\beta}p^{\alpha}p^{\beta}=\mu^2 $ now
reads:
\begin{equation}\label{13}
E^2\Delta(r)^{-1}-(\dot{r})^2\Delta(r)^{-1}-\frac{L^2}{r^2}-\mu^2=0
\end{equation}
 Solving for $\dot{r}$ we equivalently have
$ (\dot{r})^2=E^2-\Delta(r)\left(\mu^2+L^2/r^2\right)= E^2-V^2(r) $
 The  effective potential $V(r)$ is defined by the following formula
$$
\frac{V(r)}{\mu}=\sqrt{\Delta(r)\left(1+\frac{L^2}{\mu^2r^2}\right)}
 $$
and it identifies  the value of $ E/\mu$  at which the (radial) kinetic energy of the particle
vanishes.  Circular orbits correspond to  the extrema of the
effective potential, therefore solving with respect to $L$ the equation  $
\partial_{r}V=0$  we are able to find the angular momentum $L/\mu$ and then the
energy $E/\mu$   of the particle in a given circular orbit.  We
have:
\begin{equation}\label{16}
\frac{E}{\mu}=\frac{(r-2M)}{\sqrt{r(r-3M)}};\quad
\frac{L}{\mu}=\sqrt{\frac{r^2 M}{(r-3M)}},
\end{equation}
where for $r\rightarrow3M$,  $E\rightarrow\infty$. On the other hand,
solving   the circular orbits radius $r/M$ as function  of the
angular momentum we find  the following two solutions
\begin{equation}\label{17}
 \frac{r_{\pm}}{M}=\frac{L^2\pm\sqrt{L^2(L^2-12M^2\mu^2)}}{2M^2\mu^2}
\end{equation}
 The minimum radius for a stable circular orbit, $r_{\ti{lsco}}$,
occurs at the inflection points of the effective potential function,
therefore we must solve the equation $ \frac{\partial^2
V}{\partial^2 r}=0 $ (see  for example \cite{RuRR}, for a
generalization to Kerr Newmann metric \cite{Johnston:1974pn}). In
this case $r_{\ti{lsco}}=6M$:  stable circular orbits do not exist
at radii smaller than $r_{\ti{lsco}}$. Unstable circular orbits are
restricted to the range $3M < r < 6M$.
\section{Generalized Schwarzschild solution}\label{sec:GSS}
Given the 5D Ricci tensor associated to the 5D metric according to
(\ref{zippo}), we have the KK equation in vacuum : $
\ap{5}R_{\ti{AB}}=0$. The Generalized Schwarzschild Solution (GSS)
(\cite{Gross:1983hb},\cite{sorkin},\cite{davisonowen}) is a
stationary, free-electromagnetic solution  of 5D-KK equation in
vacuum,  with 4D-spherical symmetry\footnote{The ordinary
4D-spacetime $M^{\ti{4}}$ of the direct product $M^{\ti{4}}\otimes
\mathcal{S}^{\ti{1}}$ is spherically symmetric; in other words the
sections $t = cost$, $r = cost$ and $x^{\ti{5}} = cost$ of
$M^{\ti{5}}$ are $S^2$ (spherical surfaces in the ordinary
3D-space).}. After the KK reduction procedure we have equivalently a
system of 4D  Einstein equation coupled to a massless scalar field:
\begin{eqnarray}\label{ki}
G^{\mu\nu}&=&\frac{1}{\phi}\left(\nabla^{\mu}\partial^{\nu}\phi\right),\quad
\Box\phi=0\, .
\end{eqnarray}
Here $\nabla^{\mu}$ is the covariant derivative compatible with the
4D-spacetime metric and $\Box\equiv\nabla^{\mu}\nabla_{\mu}$.
\begin{figure}[h!]
\centering
\begin{tabular}{c}
\includegraphics [scale=.5]{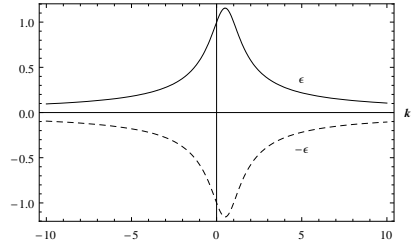}
\end{tabular}
\caption[font={footnotesize,it}]{\footnotesize{This figure
illustrates  the  $\epsilon$-parameter of the GSS
(\ref{CGMriunone}), as function of k-parameter. Solutions
$\epsilon_{\pm}$, of $\epsilon^{2}\left(k^{2}-k+1\right)=1$}, are
plotted. Every  points set a metric of the family solutions.
}\label{Dulcinea}
\end{figure}
We look for a solution of the form
\begin{equation}
ds_{\ti{(5)}}^{2}=A(\rho)dt^{2}-B(\rho)\left[d\rho^{2}+\rho^{2}d\Omega^{2}\right]-C(\rho)
dx^{5\ 2}\,.
\end{equation}
Requiring the 3D spherical symmetry and the independence of the
metric coefficients from the time $t$, we are able to obtain a family
of  exact solutions of the field equations which are
asymptotically flat:
\begin{equation}\label{INRO}
ds_{\ti{(5)}}^{2}=\left(\frac{a\rho-1}{ar+1}\right)^{2\epsilon
k}dt^{2}-\frac{1}{(a\rho)^2}
\frac{\left(a\rho+1\right)^{2\left[\epsilon(k-1)+1\right]}}{\left(a\rho-1\right)
^{2\left[\epsilon(k-1)-1\right]}}\left[d\rho^{2}+\rho^{2}d\Omega^{2}\right]-
\left(\frac{a\rho+1}{a\rho-1}\right)^{2\epsilon}dx^{5\ 2}\, ,
\end{equation}
The GSS solution is not unique, because in our framework the
Birkhoff theorem does not hold \cite{sorkin}, and it depends on
 the real parameter $\left(\epsilon,k\right)$, which are constant
 (Fig.\ref{Dulcinea})
and
constrained by
$ \epsilon^{2}\left(k^{2}-k+1\right)=1 $ .  The constant parameter $a$
is  related to the mass of a central body which is supposed to act as source. The
Schwarzschild limit is recovered for $\epsilon\rightarrow0,\;
k\rightarrow\infty$; in  such a limit $a=2c^2/(G M_{\ti{S}})$ -being
$M_{\ti{S}}$ the Schwarzschild mass and $G$  the usual Newton constant- and the above expression turns into the usual 4D  exterior solution related to a central body.
Noticeably, the Schwarzschild limit is obtained when $\phi=1$. Since
we study the exterior solution  we are able to perform the
following transformation:
\begin{equation}\label{uio}
r=\rho\left(1-\frac{r_{g}}{\rho}\right)^{2}\, .
\end{equation}
From Eqs. \ref{INRO} and \ref{uio} a  one-parameter family  is
recovered in the 4D-spherical polar coordinate\footnote{Consider
$t\in\Re$, $r\in\left]2M,+\infty\right]\subset\Re^{+}$,
$\vartheta\in\left[0,\pi\right]$, $\varphi\in\left[0,2 \pi\right]$}
$\left\{t,r,\theta,\varphi\right\}$ where
$d\Omega^2\equiv\sin^2\theta d\varphi^{2}+d\theta^{2}$. We have:
\begin{equation}\label{CGMriunone}
ds_{\ti{(5)}}^{2}=\Delta(r)^{\epsilon k}dt^{2}-\Delta(r)^{-\epsilon
(k-1)}dr^{2}-r^{2}\Delta(r)^{1-\epsilon (k-1)}d\Omega^{2}-
\Delta(r)^{-\epsilon }dx^{5\ 2}\, ,
\end{equation}
where $\Delta(r)=\left(1-2M/r\right)$. It is generally used to
explore the region $k\geq0$ and $\epsilon\geq0$ to investigate the
physical properties of
 solutions\footnote{In the cited reference and in \cite{Overduin:1998pn},
  for example, is showed how positive
  density of solution requires $k>0$ and  for
positive mass (as measured at infinity) one must have $\epsilon k >
0$.}(\ref{CGMriunone}).
 Within such a range, the GSS solution presents a naked
singularity behaviour that resolves\footnote{The  event horizon,
defined in general coordinates as the surface where the norm of the
time-like Killing vector is zero, should be located for the metric
family (\ref{INRO}) and for $\epsilon>0$ and $k>0$, in  $\rho = 1/a
$. Nevertheless the center of the 3-geometry is just at $\rho = 1/a
$. But in this point the surface area of 2-shells should goes to
zero moreover the 5D- Kretschmann scalar and the square of the
4D-Ricci tensor are divergent . Therefore  the event horizon shrinks
to the singularity  in $\rho = 1/a $. These kinds of Kaluza Klein
solitons are classified as naked singularities .} in a black hole
one only in the Schwarzschild limit for $(\epsilon,k)$.
 Here we consider no negative metric
parameters, analysing  particle motion in the region $r>2M$, (for a
review see \cite{Overduin:1998pn}). The physical meaning of the
metric parameter $k$, or alternatively the $\epsilon$ parameter, has
been widely discussed in literature. It is important to note here
that this parameter characterises the spacetime external to any
astrophysical object described by the selected GSS solution: each
values of the k-constant sets a metric solution as well as the
source associated to that solution. Following this interpretation,
there should be one different value of the $k$ parameter associated
to one different source.  In particular, in a more stringent way in
\cite{PoncedeLeon:2006xs,PoncedeLeon:2007bm}, the free metric
parameter is totally determined by measurements in 4D  by taking
into account the surface gravitational potential of the
astrophysical objects, like the Sun or other stars. Indeed, despite
of the naked singularity feature showed far from the Schwarzschild
limit, GSS solutions are supposed to describe in principle the
exterior spacetimes of any astrophysical sources that satisfy the
required metric symmetries and the source is supposed to be embedded
in a cloud provided by the scalar field. Generally, to test the
validity of such models, many efforts have been made to describe the
solar system by a GSS solution; a particular value of k, adapted to
fit the prediction of the standard gravity tests with experimental
data, has been associated to the Sun (see for a review
\cite{Overduin:1998pn}, see also \cite{Xu:2007dc}). Each comparison
gives a peculiar estimation for $k$; all these different estimations
are based on different tests assumed to probe the model validity.
Modelling the Sun by a GSS solution should require a fine tuning of
the characteristic parameter. For example, in
\cite{Overduin:2000gr}, experimental constraints on equivalence
principle violation in the solar system translate in a $k>5.\times
10^7$. Extra dimensions play thus a negligible role in the solar
system dynamics. Meanwhile, by measures of the surface gravitational
potential in \cite{PoncedeLeon:2006xs,PoncedeLeon:2007bm}, the Sun
seems to be characterized by a $k=2.12$. On the other hand  all the
standard tests on light-bending around the Sun, or the perihelion
precession of Mercury, constrain $k \gtrsim 14$.

\section{Time-like circular orbits in the GSS
spacetimes.}\label{sec:CircularorGSS}
\subsection{Geodesic approach}
Studying the timelike circular orbits in the background
(\ref{CGMriunone}), at  first we consider  test particles motion in
the geodesic approach where a constant test particle mass
$\mu_{\ti{(5)}}=cost$ is considered. Let us consider the  5D
momentum
$\ap{5}P^{\ti{A}}\equiv\mu_{\ti{(5)}}\omega^{\ti{A}}=\mu_{\ti{(5)}}\alpha
u^{\ti{A}}$ or, equivalently, $\ap{5}P^{\ti{A}}\equiv\alpha
\ap{4}P^{\ti{A}} $, where
$\ap{4}P^{\ti{A}}\equiv\mu_{\ti{(5)}}u^{\ti{A}}$ and
$\ap{5}P^{\ti{A}}\equiv\mu_{\ti{(5)}}\omega^{\ti{A}}$. Granted the
three killing vectors
\begin{equation}
\xi^{\ti{A}}_{(t)}\equiv\{1,0,0,0,0\},\quad \
\xi^{\ti{A}}_{(5)}\equiv\{0,0,0,0,1\},\quad\xi^{\ti{A}}_{(\varphi)}\equiv\{0,0,0,1,0\},
\end{equation}
the following conserved quantities in $(M^{\ti{5}},g^{\ti{5}})$ can
be defined:
\begin{equation}\label{rip}
\ap{5}\mathcal{E}\equiv\xi^{\ti{A}}_{(t)}\ap{5}P_{\ti{A}} ,\quad \
\ap{5}\Gamma\equiv\xi^{\ti{A}}_{(5)}P_{\ti{A}},\quad \ap{5}L\equiv
\xi^{\ti{A}}_{(\varphi)}\ap{5}P_{\ti{A}}.
\end{equation}
 Introducing the quantities
$\ap{4}\mathcal{E}\equiv\xi^{\ti{A}}_{(t)}\ap{4}P_{\ti{A}}$ and
$\ap{4}\mathcal{L}\equiv\xi^{\ti{A}}_{(\varphi)}\ap{4}P_{\ti{A}}$,
that are in general non constant along the motion, we can also
write: $\ap{5}\mathcal{E}=\alpha\ap{4}\mathcal{E}$, $\ap{5}L=\alpha
\ap{4}L$ and $\ap{5}\Gamma=\mu_{\ti{(5)}}g_{55}
\omega^{5}=\mu_{\ti{(5)}}\alpha g_{55} u^{5} $. For neutral
particles $\ap{5}\mathcal{E}=\ap{4}\mathcal{E}$ and $
\ap{5}L=\ap{4}L$.
 The Schwarzschild limit of the constants of motion is
\begin{equation}\label{crepuscolo}
\ap{5}\mathcal{E}=\left(1-u_{5}^{2}\right)^{-1/2}\ap{4}\mathcal{E}\quad\mbox{,}\ap{5}L=
\left(1-u_{5}^{2}\right)^{-1/2}
\ap{4}L\, .
\end{equation}
In this limit $\ap{5}\mathcal{E}=\ap{4}\mathcal{E}$ and $\ap{5}L=
\ap{4}L$ on the surfaces $x^5=cost$, where Eqs.\ref{Revisione} are
geodesic; hence $\ap{4}\mathcal{E}$ and $\ap{4}\mathcal{L}$ are
respectively interpreted as the energy at infinity and the total
angular momentum.  In general, in the case of charged particle, we
interpret $\ap{5}\mathcal{E}$ and $\ap{5}L$. as the energy at
infinity and
 the total angular momentum  of a particle following the
trajectory  defined by the first of (\ref{Revisione}) in
$\left(M^{\ti{4}}, g^{\ti{4}}\right)$. A 5D-Lagrangian
$\ap{5}\mathcal{L}$ and its 4D-counterpart $\ap{4}\mathcal{L}$ are
defined  as $\ap{5}\mathcal{L}\equiv g_{\ti{AB}}\ap{5}P^{\ti{A}}
\ap{5}P^{\ti{B}}$, $\ap{4}\mathcal{L}\equiv
g_{\mu\nu}\ap{4}P^{\mu}\ap{4}P^{\nu} $. Explicitly we have:
$\ap{5}\mathcal{L}=\alpha^{2}
\ap{4}\mathcal{L}+\ap{5}\Gamma^{2}/g_{\ti{55}}\equiv\alpha^{2}
\mu_{\ti{(5)}}^{2}
g_{\ti{rr}}(\dot{r})^{2}+\ap{5}\mathcal{E}^{2}/g_{\ti{00}}
+\ap{5}L^{2}/g_{\varphi\varphi}+\ap{5}\Gamma^{2}/g_{\ti{55}}$ where
$\dot{r}\equiv u^{r}$. An effective potential is usually defined via
the value of $\ap{5}\mathcal{E}/\mu_{\ti{(5)}}$ at which the
(radial) kinetic energy of the particle vanishes. We have:
\begin{equation}
 \ ^{\ti{(5)}}\!V_{eff}\equiv\sqrt{ g_{\ti{00}} \left[1-\left( \frac{ \
^{\ti{(5)}}\!L^{2}}{\mu_{\ti{(5)}}^{2}g_{\varphi\varphi}}+ \frac{ \
^{\ti{(5)}}\!\Gamma^{2}}{\mu_{\ti{(5)}}^{2}g_{\ti{55}}}\right)\right]},\quad
\mbox{or}\quad  \ ^{\ti{(5)}}\!V_{eff}\equiv\sqrt{g_{\ti{00}}
\left(\alpha^{2}- \frac{ \
^{\ti{(5)}}\!L^{2}}{\mu_{\ti{(5)}}^{2}g_{\varphi\varphi}}\right)}\,
.
\end{equation}
More explicitly:
\begin{equation}
\ap{5}V_{eff}\equiv\sqrt{\Delta^{\epsilon k}
\left[1+r^{2}\Delta^{-1+\epsilon
(k-1)}\frac{\ap{5}L^{2}}{\mu_{\ti{(5)}}^{2}}+ \Delta^{\epsilon
}\frac{\ap{5}\Gamma^{2}}{\mu_{\ti{(5)}}^{2}} \right]}
\end{equation}
The condition for the occurrence of the circular orbits is: $
\partial\ap{5}V_{eff}/\partial r=0 $.  Solving this equation with respect to
 $\ap{5}L$  we find
and the angular momentum $\ap{5}L/(M \mu_{\ti{(5)}})$ and the energy
$\ap{5}\mathcal{E}/\mu_{\ti{(5)}}$ of a particle in a circular orbit
of radius $r$:  $$ \ap{5}\mathcal{E}=\mu_{\ti{(5)}}
\sqrt{\left[\left(\frac{g_{\varphi\varphi}}{g_{00}}\right)_{,r}\right]^{-1}\left[g_{\varphi\varphi
,r}-\frac{\ap{5}\Gamma^{2}}{\mu_{\ti{(5)}}^{2}}\left(\frac{g_{\varphi\varphi}}{g_{55}}\right)_{,r}
\right]}$$.
 $$ \ap{5}L^{\pm}=\pm \mu_{\ti{(5)}}
\sqrt{\left[\left(\frac{g_{00}}{g_{\varphi\varphi}}\right)_{,r}\right]^{-1}\left[g_{00
,r}-\frac{\ap{5}\Gamma^{2}}{\mu_{\ti{(5)}}^{2}}\left(\frac{g_{00}}{g_{55}}\right)_{,r}
\right]} $$ In terms of $\ap{4}\mathcal{E}$ and $\ap{4}\mathcal{L}$
we have:
$$
\frac{\ap{5}\mathcal{E}}{\mu_{\ti{(5)}}} =\sqrt{1+\Delta^{\epsilon
}\frac{\ap{5}\Gamma^{2}}{\mu_{\ti{(5)}}^2}}\frac{\ap{4}\mathcal{E}}{\mu_{\ti{(5)}}}
,\quad\frac{\ap{5}L}{\mu_{\ti{(5)}} M}=\sqrt{1+\Delta^{\epsilon }
\frac{\ap{5}\Gamma^{2}}{\mu_{\ti{(5)}}^2}}\frac{\ap{4}L}{\mu_{\ti{(5)}}
M}\, .
$$
 The Schwarzschild limit on the energy and angular momentum gives following result:
$$
\lim_{k\rightarrow\infty}\frac{\ap{5}L}{\mu_{\ti{(5)}} M}=
\frac{r}{M}
\sqrt{\frac{M}{r-3M}\left(1+\frac{\ap{5}\Gamma^{2}}{\mu_{\ti{(5)}}^2}\right)}
$$
$$
\lim_{k\rightarrow\infty}\frac{\ap{5}\mathcal{E}}{\mu_{\ti{(5)}}}=\sqrt{\Delta\left(\frac{r-2M}{r-3
M}+\frac{\ap{5}\Gamma^{2}}{\mu_{\ti{(5)}}^2}\frac{2M-r}{3M-r}\right)}
$$
As far as the circular orbit radius $r_{c}$ is concerned, we infer,  in agreement with the known result in
literature:
\begin{equation}\label{RCIC}
r_{c}>[1+\epsilon(2k-1)]M\,.
\end{equation}
The above expression  is a free-$\ap{5}\Gamma$ quantity with $
\lim_{k\rightarrow\infty}r_{c}=3M$ and $ r_{c}<3M$ $\forall k>0. $
\begin{figure}[h!]
\centering
\begin{tabular}{c}
\includegraphics[scale=.5]{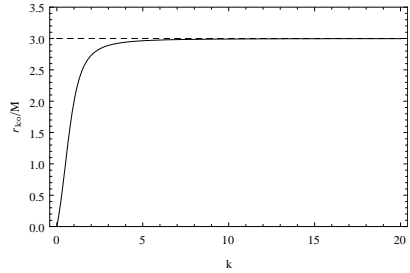}
\end{tabular}
\caption[font={footnotesize,it}]{\footnotesize{Test particle last
circular orbit radius  $r_{lco}=1+(2k-1)/\sqrt{k^2-k+1}$ is plotted
as function of the  metric parameter $k$. The Schwarzschild limit,
$r_{lco}=3M$, is also plotted (dashed line). The last circular orbit
radius approach   $r_{lco}=3M$ for large value of $k$. Otherwise
$r_{lco}<3M$: circular orbits (unstable and stable) are possible
also in a region $r<3M$.}} \label{RcEK}
\end{figure}
For an extensive analysis of the motion in the GSS background  see
\cite{Overduin:1998pn}, \cite{Liu:2000zq}-\cite{Kalligas:1994vf}.
The above result implies that the last circular orbit does not
depend on the particle charge but it is rather a geometrical feature
of the selected metric solution. Moreover, particles in circular
orbit (stable or unstable) should be detectable also at values
$r<3M$.

\subsection{Papapetrou analysis}
Let us consider now the 4D-dispersion relation (\ref{zanzibar}),
$P_{\mu}P^{\mu}=m^{2}$ where $P_{\mu}=m u_{\mu}$ .  Mass is now a
varying term, although  in this particular case  it turns out to be
a function of radial coordinate $r$ only and this means that $m$ is
constant along the circular orbits (at fixed $r$). Only in the
Schwarzschild limit, or asymptotically, where $\phi=1$ we have
$m=m_0=cost$. Anyway, it is always possible to build the  constants
of motion $\mathfrak{E}$ and $\mathfrak{L}$ defined as follows:
\begin{equation}\label{WIWA}
\mathfrak{E}=p_{0}=m g_{00} u^{0}, \quad \mathfrak{L}=p_{\varphi}= m
g_{\varphi\varphi} u^{\varphi}\, .
\end{equation}
An effective potential for a  test particle  of mass $m$  can be
defined \footnote{In this case the $\mathfrak{V}_{eff}$ has unit of
mass.}  adopting the standard procedure. We have\footnote{The
effective potential now depends of the non constant mass $m$,
 this fact could  be alternative seen as a direct dependence of the
potential by the matter field $\phi$. Anyway we remark that of
circular orbits the particle mass turns out to be a constant.}:
\begin{equation}\label{Maevero?SperiamoAltrimeti}
\mathfrak{V}_{eff}\equiv\mathfrak{E}=
\sqrt{g_{00}\left(m^{2}-\frac{\mathfrak{L}^{2}}{g_{\varphi\varphi}}\right)
}\end{equation} We now focus    on some interesting scenarios
prospected by the Papapetrou approach applied to    our analysis of
motion into the GSS background.  We consider here the case in which
the dynamical parameter $A$ is a function of spacetime point or
$A=\beta m \phi^2$, where $\beta$ is a real number. As a  particular
subcase, imposing $\beta=0$ we at first focus  on the case $A=0$.
\subsubsection{$A=0$}\label{A0}
\begin{figure}[h!]
\centering
\begin{tabular}{cc}
\includegraphics[scale=.5]{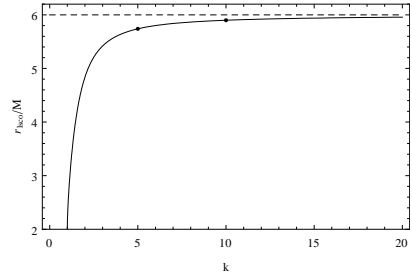}
\end{tabular}
\caption[font={footnotesize,it}]{\footnotesize{In this picture last
stable circular orbit radius $r_{\ti{lsco}}/M$, obtained from the
Papapetrou's approach with $A=0$ (see (\ref{cam})), is plotted as
function of the  $k$-parameter. Schwarzschild limit, $r_{lsco}=6M$,
is also plotted (dashed line). Last stable circular orbit radius
approach to the $6M$ for large value of $k$. Otherwise
$r_{\ti{lsco}}<6M$: stable circular orbits are possible also in a
region $r<6M$.}} \label{PlotRlsco}
\end{figure}

Equations of motion (\ref{padredimaria}) when $A=0$ became
\begin{equation}\label{PapA0}
u^{a}\,^{\ti{(4)}}\nabla_{a}u^{b}=0\quad\mbox{and}\quad\partial_{\mu}m=0\,.
\end{equation}
These equations describe a geodetic motion in the ordinary
4D-spacetime for a test particle of constant mass $m$,  where no
scalar field coupling term appears. Formally, these are the same
equations of motions we have in (\ref{Revisione}), where
$\omega_{5}=0$ and $\mu_{\ti{(5)}}=m$. Nevertheless,
Eq.(\ref{PapA0})
 describes charged as well as neutral particles.     The following conserved
quantities  in $(M^{\ti{4}},g^{\ti{4}})$ can be defined:
$\mathcal{E}_{\ti{$\epsilon k$}}\equiv\xi^{a}_{(t)}P_{a}$,
$L_{\ti{$\epsilon k$}}\equiv \xi^{a}_{(\varphi)}P_{a}$.
Equivalently, we have:
\begin{equation}
\mathcal{E}_{\ti{$\epsilon k$}}=m\Delta^{k \epsilon }\dot{t},
L_{\ti{$\epsilon k$}}=-m\dot{\varphi} r^{2}\Delta^{(1-k) \epsilon
+1}\csc(\theta)^{-2}.
\end{equation}
As usually we can define the quantities $\mathcal{E}$ and $L$ such that
\begin{equation}\label{11}
\mathcal{E}_{\ti{$\epsilon k$}}\equiv\mathcal{E}\Delta^{\epsilon k
-1}, \quad\ L_{\ti{$\epsilon k$}}\equiv L\Delta^{\epsilon(1-k)+1}.
\end{equation}
In the Schwarzschild limit $\mathcal{E}_{\ti{$\epsilon
k$}}=\mathcal{E}$ and  $\mathcal{L}_{\ti{$\epsilon k$}}=\mathcal{L}
$; we interpret the (\ref{11}) as  the energy at infinity and
 the total angular momentum  of the particle.
A Lagrangian  $\mathcal{L}_{\ti{$\epsilon k$}}$ is defined as follows:
$$
\mathcal{L}_{\ti{$\epsilon k$}}\equiv-m^{2}\Delta^{-\epsilon(k-1)}
\left(\dot{r}\right)^{2}+\mathcal{E}_{\ti{$\epsilon
k$}}^{2}\Delta^{-\epsilon k}-\frac{L_{\ti{$\epsilon
k$}}^{2}}{r^{2}}\Delta^{\epsilon(k-1)-1}\,.
$$
The effective potential $V_{\ti{$\epsilon k$}}\equiv E/m$ is defined
in the usual way as
$$
V_{\ti{$\epsilon k$}}\equiv\sqrt{ g_{\ti{00}} \left(1-
\frac{L_{\ti{$\epsilon k$}}^{2}}{m^{2}g_{\varphi\varphi}}\right)}\,.
$$
The energy $\mathcal{E}_{\ti{$\epsilon k$}} $ and the angular
momentum   $L_{\ti{$\epsilon k$}} $ of a massive test particle in a
circular orbit are
\begin{equation} \label{seduta}
\mathcal{E}_{\ti{$\epsilon k$}} =m \sqrt{g_{\varphi\varphi
,r}\left[\left(\frac{g_{\varphi\varphi}}{g_{00}}\right)_{,r}\right]^{-1}}
,\quad L^{\pm}_{\ti{$\epsilon k$}}=\pm m \sqrt{g_{00
,r}\left[\left(\frac{g_{00}}{g_{\varphi\varphi}}\right)_{,r}\right]^{-1}}\,,
\end{equation}
where  the Schwarzschild limit provide the following free
$\Gamma_{\ti{5}}$-quantities:
$$
\frac{L_{\ti{$\epsilon k$}}}{M m}=\frac{r}{M}\sqrt{
\frac{1}{\frac{r}{M}-3}},\quad \frac{\mathcal{E}}{m}=\sqrt{\Delta
\frac{r-2M}{r-3 M}}\,.
$$
 From (\ref{seduta}) we infer $ r_{c}>\left[1+\epsilon(2k-1)\right]M
$ for the circular orbits radius $r_{c}$ (Cfr.\ref{RCIC}). The
turning points of the effective potential are located in
\begin{equation}\label{cam}
r^{\pm}=\left[1+\epsilon(3k-2)\pm\epsilon\sqrt{(-1+k) (-1+4
k)}\right]M\,,
\end{equation}
where last stable circular orbit radius is $r^+=r_{\ti{lsco}}$,
while in the Schwarzschild limit $r^{+}\equiv6M$ ad $r^{-}\equiv2M$.
Moreover it is possible to see, Fig.\ref{PlotRlsco}, that
$r_{\ti{lsco}}<6M$, $\forall k>0$. It is worth noting that the last
stable circular orbit radius is located below its Schwarzschild
limit; this means that in principle there could be particles in
stable orbits for values of radius orbit just less that $6M$, and
this  represents a valid constraint in order to compare theory with
experimental data. In Tab\ref{tab:grossecic} we report the range
$|r_{\ti{lsco}}-6M|$, for selected\footnote{Others constraints of
the metric parameters, based on the motion analysis are given in
\cite{Overduin:1998pn}, \cite{Liu:2000zq},
\cite{PoncedeLeon:2006xs}, \cite{Liko:2003ha}, \cite{T4},
\cite{Kalligas:1994vf}. } values of $k$.
The energy and angular momentum of the last circular orbits are
:
 \begin{equation}\label{Dire}
\frac{ L^{\pm}_{\ti{$\epsilon k$}}}{Mm}=\pm
\frac{r^{+}_{\ti{lsco}}}{M} \left(1-\frac{2
M}{r^{+}_{\ti{lsco}}}\right)^{\frac{1}{2}\left[(1-k)\epsilon+1\right]}\sqrt{\frac{\epsilon
k }{\epsilon\left(1-2
k\right)+\left(\frac{r^{+}_{\ti{lsco}}}{M}-1\right)}}\,,
\end{equation}
\begin{equation}\label{Straits}
\frac{\mathcal{E}_{\ti{$\epsilon k$}} }{m}=\left(1-\frac{2
M}{r^{+}_{\ti{lsco}}}\right)^{\frac{k\epsilon}{2}}
\sqrt{1-\frac{\epsilon k}{\epsilon(2k-1)+\left(1 -\frac{
r^{+}_{\ti{lsco}}}{M}\right)} }\,.
\end{equation}
\begin{figure}[h!]
\centering
\begin{tabular}{cc}
\includegraphics[scale=.5]{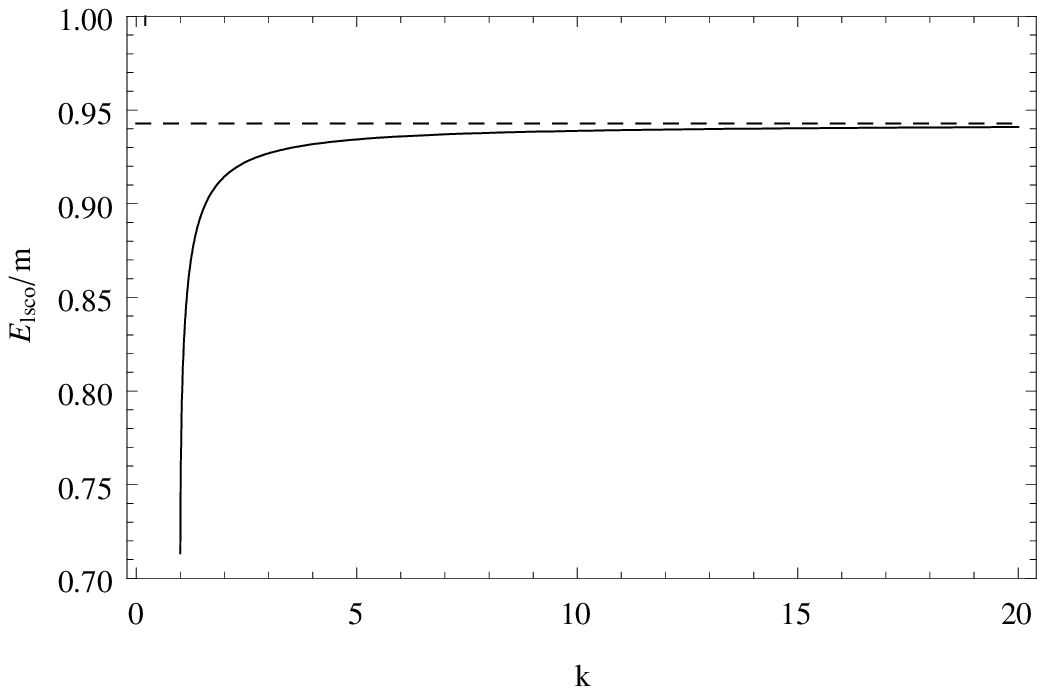}&
\includegraphics[scale=.5]{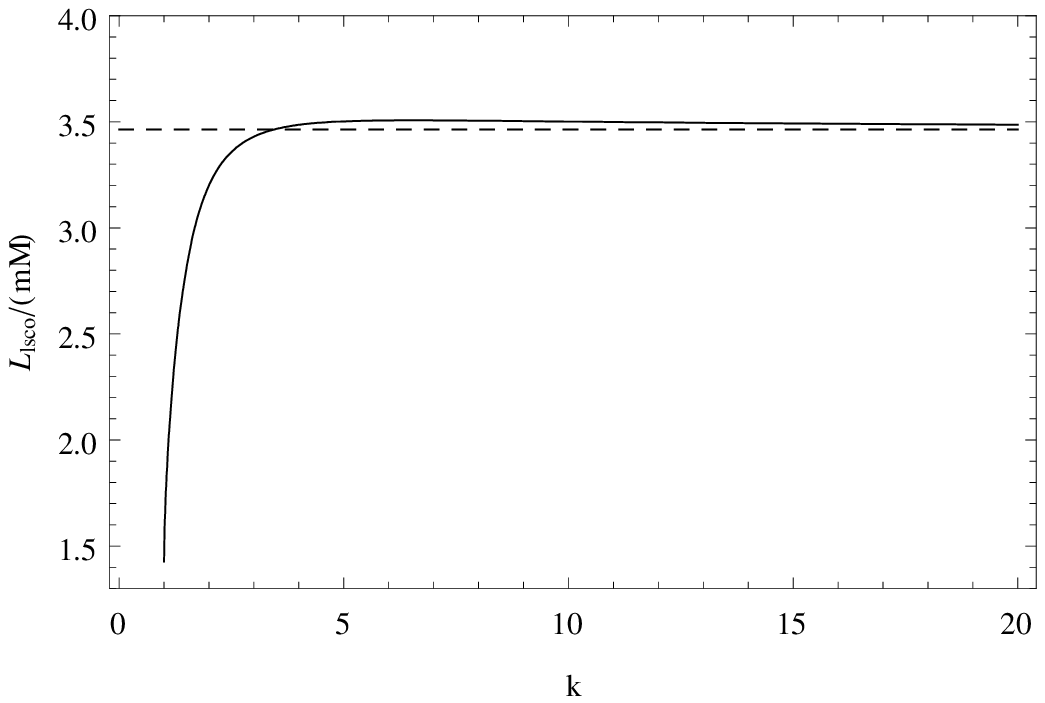}\\(a)&(b)\\
\end{tabular}
\caption[font={footnotesize,it}]{\footnotesize{The  ``energy''
$E_{\ti{lsco}}/m_0$ and the ``angular momentum'' $L_{\ti{lsco}}/m_0$
in the circular orbits, obtained by Papapetrou's approach with $A=0$
(see (\ref{Dire}) and (\ref{Straits})), are plotted as functions
k-parameter. Schwarzschild' limits for the energy and the angular
momentum are also plotted (dashed lines). The energy
$\mathcal{E}_{\ti{lsco}}$ is always below its Schwarzschild limit
while $\mathcal{L}_{\ti{lsco}}$, for $k>3.45644$ is over the
Schwarzschild limit.}} \label{PLOTLEK}
\end{figure}
It is possible to see, Fig.\ref{PLOTLEK}, that the energy
$\mathcal{E}_{\ti{$\epsilon k$}}$ for all values of $k$-parameter is
always below its Schwarzschild limit, the angular momentum
$\mathcal{L}_{\ti{$\epsilon k$}}$ is beyond the Schwarzschild limit
for $k>3.45644$. This fact should not be read as a direct
consequence of a possible motion along the fifth dimension, since
Eq.(\ref{PapA0}) does not depend on it, neither on the
$g_{\ti{55}}$-metric component. We interpret it as a features
related to deformation of the Schwarzschild metric as long as $k$ is
sufficiently small; see also Eq.(\ref{seduta}). This  seems to be
confirmed also by the fact that Eqs.(\ref{Dire}, \ref{Straits}) are
the same that one can obtain from the geodesic approach with
$\omega_{5}=0$.

 In the
following analysis we choice different values of the dynamical
parameter $A$ where $\mathcal{E}_{\ti{lsco}}$ and
$\mathcal{L}_{\ti{lsco}}$ have the same behaviour.
\subsection{$A=cost$}
\begin{figure}[h!]
\centering
\begin{tabular}{cc}
\includegraphics[scale=.5]{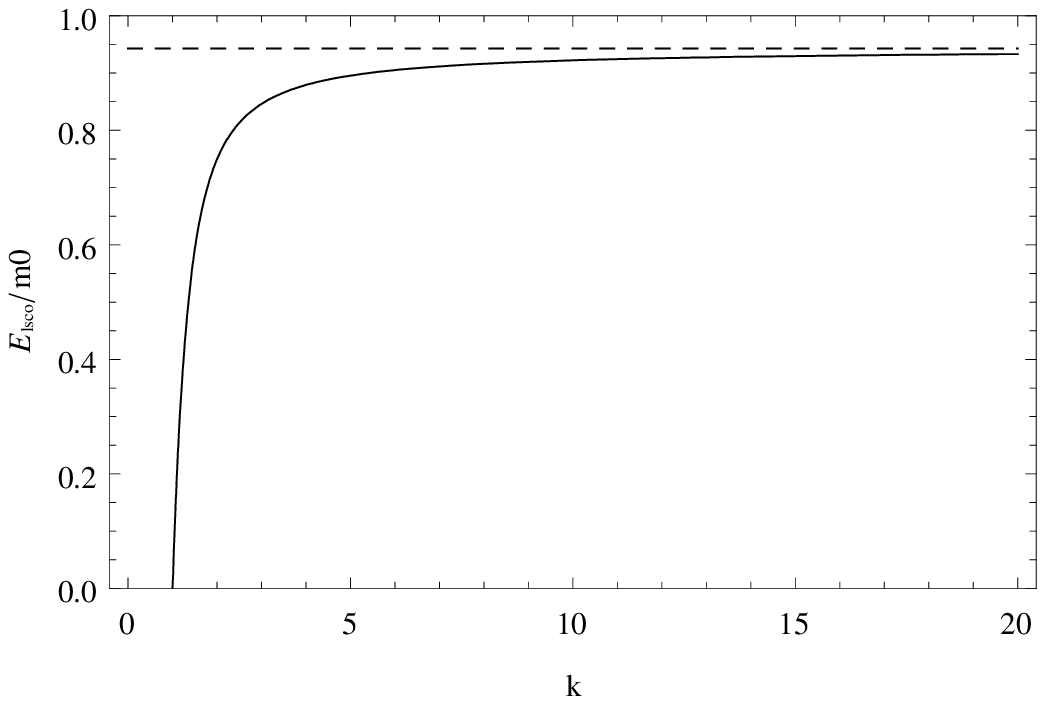}&
\includegraphics[scale=.5]{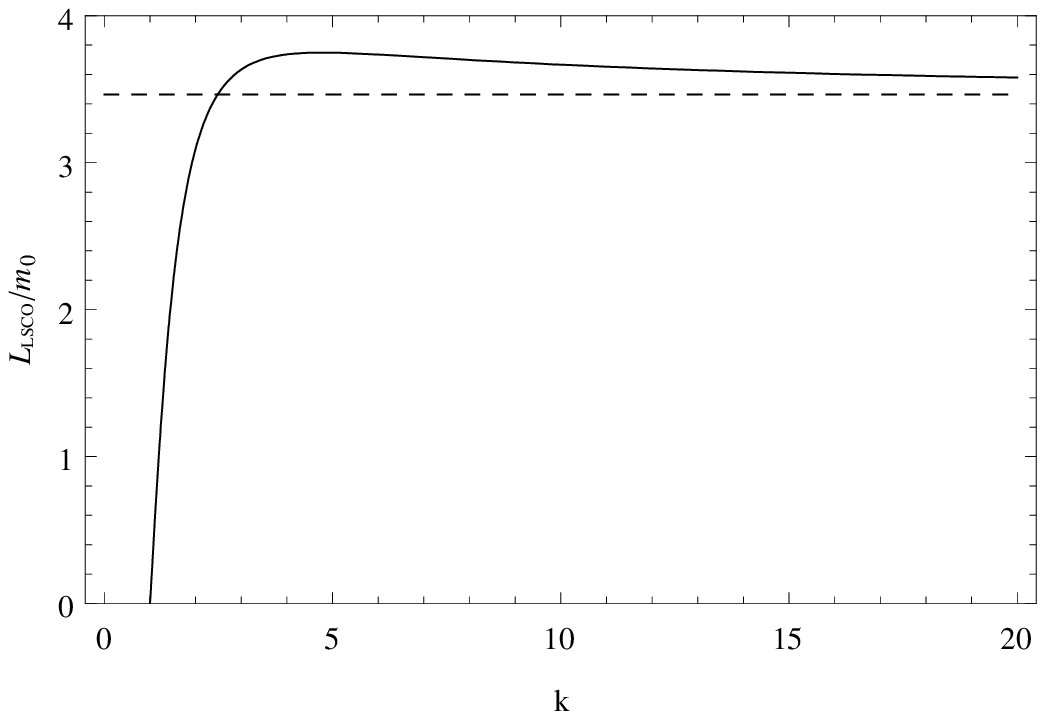}\\(a)&(b)\\
\end{tabular}
\caption[font={footnotesize,it}]{\footnotesize{The energy
$E_{\ti{lsco}}/m_0$ and angular momentum $L_{\ti{lsco}}/m_0$ of
circular orbits, obtained by Papapetrou's approach with $A=cost$,
see (\ref{cortileE}) and (\ref{cortileL}), are plotted as
k-parameter functions. Schwarzschild limits for the energy and the
angular momentum are also plotted (dashed lines). The energy
$\mathcal{E}_{\ti{lsco}}$ is always under its Schwarzschild limit
while $\mathcal{L}_{\ti{lsco}}$, for $k>2.48491$ is over the
Schwarzschild limit.}} \label{PLOTLEKboh}
\end{figure}

As a simplest generalization of the previous case we are going to
consider  $A=cost$. Integrating  Eq.(\ref{padredisofia}) along a
curve $\gamma=\gamma(s)$, between the points $P=\gamma(s)$ and
$P_0=\gamma(s_0)$, we obtain:
\begin{equation}\label{Bachsinfonia}
m=\frac{A}{2\phi^{2}}+m_{0}-\frac{A}{2\phi_{0}^{2}}\,.
\end{equation}
In the Schwarzschild limit $m=m_{0}$; we set\footnote{The
dynamical parameter $A$ is here related to the initials conditions
of the particle motion.}   $A= 2 m_0\phi_{0}^{2} $, therefore $m= A/2\phi^2$.
Eq.(\ref{padredimaria}) becomes now
\begin{equation}\label{padredimaria2}
 u^{a}\ ^{\ti{(4)}}\!\nabla_{a}u^{b}=
(u^{b}u^{c}-g^{bc})\left(2\frac{\partial_{c}\phi}{\phi}\right)\,,
\end{equation}
which  does not depend on $A$. The effective
potential (\ref{Maevero?SperiamoAltrimeti}) in this case reads:
\begin{equation}\label{MaeveroAcos}
\mathfrak{V}_{eff}= \sqrt{g_{00}\left(\frac{A^{2}}{4
\phi^{4}}-\frac{\mathfrak{L}^{2}}{g_{\varphi\varphi}}\right)
}\,.
\end{equation}
The  momentum $\mathfrak{L}$ and the energy $\mathfrak{E}$
for  circular time-like orbits are :
\begin{equation}\label{cortileL}
\mathfrak{L}^{2}=\frac{A^{2}}{4\phi^{3}}\left[\frac{d}{dr}\left(\frac{g_{00}}{g_{\varphi\varphi}}\right)\right]^{-1}
\left[\frac{d}{dr}\left(\frac{g_{00}}{\phi}\right)-\frac{3g_{00}}{\phi^2}\frac{d\phi}{dr}\right]
\end{equation}
and
\begin{equation}\label{cortileE}
\mathfrak{E}=\sqrt{\frac{A^{2}}{4\phi^{3}}\left[\frac{d}{dr}\left(\frac{g_{\varphi\varphi}}{g_{00}}\right)\right]^{-1}
\left[\frac{d}{dr}\left(\frac{g_{\varphi\varphi}}{\phi}\right)-\frac{3g_{\varphi\varphi}}{\phi^2}\frac{d\phi}{dr}\right]
}\,.
\end{equation}
In the Schwarzschild limit:
\begin{equation}\label{cortileL0lim}
\mathfrak{L}^{2}=\left(\frac{m_{0} r\sqrt{M }}{2\sqrt{-3
M+r}}\right)^{2} \quad\mbox{and}\quad \mathfrak{E}=\frac{1}{2}
\sqrt{-\frac{m_{0}^2(-2 M+r)^2}{(3 M-r) r}}\,.
\end{equation}
 Last circular orbit  is located at $r_{\ti{lco}}\equiv M
 \left[1+\epsilon\left(2k -1\right)\right] $
- Cfrt.Eq.(\ref{RCIC}).
For the last stable circular orbit, as the turning
point of the effective potential, (\ref{MaeveroAcos}) we have:
\begin{equation}\label{Formia20090}
r_{\ti{lsco}}\equiv\frac{\sqrt{M^{2}\left[4+(15k-8)
\epsilon^{2}+5(8-3k)\epsilon
^{4}\right]}+M\left[3+\epsilon(2+k-11\epsilon+5k\epsilon)\right]}{(2+k)\epsilon}\,.
\end{equation}
\begin{figure}[h!]
\centering
\begin{tabular}{cc}
\includegraphics[scale=.5]{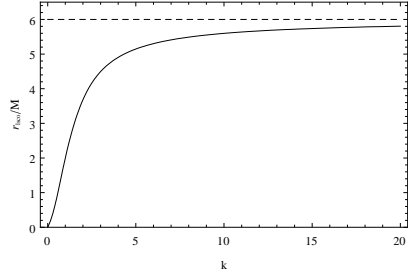}
\end{tabular}
\caption[font={footnotesize,it}]{\footnotesize{In this picture last
circular orbit radius $r_{\ti{lsco}}/M$, obtained from the
Papapetrou's approach with $A=cost$ (see (\ref{Formia20090})), is
plotted as function of the  $k$-parameter. Schwarzschild limit,
$r_{lsco}=6M$, is also plotted (dashed line). Last stable circular
orbit radius approach to the $6M$ for large value of $k$. Otherwise
$r_{\ti{lsco}}<6M$: stable circular orbits are possible also in a
region $r<6M$. }} \label{PLOTLEKboh0}
\end{figure}
We have a free-A quantity, but it is a function of the only metric
parameters $(\epsilon, k)$. Also in this case $r_{\ti{lsco}}<6M$ and
in the Schwarzschild limit $r_{\ti{lsco}}=6M$. See also
Tab\ref{tab:grossecic}. The energy $\mathfrak{E}_{\ti{lsco}}/m_{0}$
and the  momentum $\mathfrak{L}_{\ti{lsco}}/(m_{0}M)$ in the last
stable circular orbits are plotted \footnote{In the cases $A=cost$,
and $A=\beta m \phi^{2}$ these are not considered as functions of
$m=m(r_{\ti{lsco}})$ but only $m_0$. The comparison with case of
$E_{\ti{lsco}}/m_{\ti{lsco}}$ will be detailed discussed in another
work \cite{Mio1}.} in Fig.\ref{PLOTLEKboh}.
\subsubsection{$A=\beta m \phi^{2}$}
In the case $A=\beta m \phi^{2}$, where $\beta$ is a real number,
the mass $m$ is no more a constant, but integrating along a curve
$\gamma=\gamma(s)$, between the points $P=\gamma(s)$ and
$P_0=\gamma(s_0)$, the following scaling law arises:
\begin{equation}\label{galleriadopoformia}
m=\frac{m_{0}\phi_{0}^{\beta}}{\phi^{\beta}}\,.
\end{equation}
Here $m_{0}\phi_{0}^{\beta}=cost$ and the equations of motion
becomes
\begin{equation}\label{Rachmaninovdopo}
u^{a}\ ^{\ti{(4)}}\!\nabla_{a}u^{b}=
\left(u^{b}u^{c}-g^{bc}\right)\frac{\partial_{c}\phi}{ \phi}\beta\,,
\end{equation}
therefore it does not depend on $m$ but on the constant $\beta$.
The present case   reduces to the $A=cost$-case when one
sets $\beta=2$ and  $A^{2}=4 m_{0}^{2} \phi_{0}^{2\beta}$.
Introducing the parameter $B^{2}\equiv m_{0}^{2} \phi_{0}^{2\beta}$,
the effective potential reads:
\begin{equation}\label{MaeveroAnocos}
\mathfrak{V}_{eff}= \sqrt{g_{00}\left(\frac{B^{2}}{ \phi^{2
\beta}}-\frac{\mathfrak{L}^{2}}{g_{\varphi\varphi}}\right)
}\,.
\end{equation}
The constants  $\mathfrak{L}$ and $\mathfrak{E}$ are respectively
given by:
\begin{equation}\label{cortilenoL0}
\mathfrak{L}^{2}=\left(-1\right)^{\beta}\left(1-\frac{2
M}{r}\right)^{(\beta+1-k)\epsilon}\frac{(2M-r)(k+\beta)\epsilon
B^{2}M r}{M-r+(2k-1)M\epsilon}
\end{equation}
and
\begin{equation}\label{cortilenoE0}
\mathfrak{E}=\sqrt{\frac{B^2\Delta(r)^{(k+\beta)\epsilon}
\left(-1\right)^{\beta}
[M-r+M(-1+k-\beta)\epsilon]}{M-r+(2k-1)M\epsilon}}\,.
\end{equation}
In the Schwarzschild limit they became:
\begin{equation}\label{cortileL0lim2}
\mathfrak{L}^{2}=(-1)^{\beta}\frac{r^{2} m_0^{2}M}{r-3M},\quad
\mathfrak{E}=\sqrt{-\frac{(-1)^{\beta}m_0^2(r-2M)^2}{(3M-r)r}}\,,
\end{equation}
where $B=m_{0}$ and $\beta=2n$ with $n\in\mathbb{Z}$. In general,
for $k>-\beta$ last circular orbit is located at $ r_{lco}\equiv M
\left[1+\epsilon\left(2k +1\right)\right] $
and $r_{lco}<3M$. In the Schwarzschild limit $r_{lco}=3M$. Last
stable circular orbit radius, as the turning point of the effective
potential (\ref{MaeveroAnocos}), is in
$$
r_{\ti{lsco}}\equiv
M\frac{3+\epsilon[k+\beta+(-3+k+2k\beta-\beta(2+\beta))\epsilon]}{(k+\beta)\epsilon}
+
$$
$$
+M\frac{\sqrt{4+\epsilon^{2}\left[-3k(1+2\beta)\left(\epsilon^{2}-1\right)+(2+\beta)\left(\beta-4
+\left(\beta^{3}+2\right)\epsilon
^{2}\right)\right]}}{(k+\beta)\epsilon}\,
$$
while  in the Schwarzschild limit $r_{\ti{lsco}}=6M$. Radius of last
stable circular orbit  depends on two free parameters, $k$ ,
\textit{i.e.} the independent  metric parameter and $\beta$, namely
the  ``dynamical'' one. Moreover  $r_{\ti{lsco}}<6M$ for $\beta>0$,
while for $\beta<0$ and $k>-\beta$,  $r_{\ti{lsco}}>6M$ is possible.
For $\beta=2$ we recover the same physical situations sketched in
the case $A=0$.

 More generally it is possible to see that  at an increase of
$\beta>0$ for fixed values of the parameter $k$ provides an increase
of the difference $|r_{\ti{lsco}}-6M|$ as listed in
Tab.\ref{tab:grossecic} for selected values of $k$ and $\beta$. 
\begin{table}
\caption{Differences $|r_{lco}-3M|$ for the last circular orbit
radius, and $|r_{lsco}-6M|$ for the last stable circular orbit
radius are listed for selected values of $A$.}
\label{tab:grossecic}      
\begin{tabular}{lllllll}
\hline\noalign{\smallskip}
$k$ &$|r_{\ti{lco}}-3M|$ &$$&$$&$|r_{\ti{lsco}}-6M|\quad(M)$&\\
$$ &$(M)$&$A=0$& $A=cost$& $A=-2 m \phi^{2}$& $A=4 m \phi^{2}$\\
\noalign{\smallskip}\hline\noalign{\smallskip}
 $5$ &$0.04$ &$0.3$&$0.85$&$0.411$&$1.34$
\\
$10$ &$0.008$ &$0.1$&$0.40$&$0.21$&$0.68$
\\
$20$ &$0.002$&$0.04$&$0.19$&$0.11$&$0.34$
\\
$100$ &$10^{-4}$&$0.008$&$0.04$&$0.067$&$0.22$\\
\noalign{\smallskip}\hline
\end{tabular}
\end{table}
\section{Conclusions}\label{sec:conclusion}
The dynamic in Generalized Schwarzschild solution (GSS) spacetimes
has been explored studying an effective potential for massive test
particles in circular orbits. First we have analysed the motion by
the standard approach to the particle dynamics in Kaluza Klein,
therefore considering 5D-particles moving along geodesic curves in a
5D-manifold. Then we consider the motion by an approach  \emph{a
l$\acute{a}$} Papapetrou to the dynamic considering a 5D particle
described by an energy momentum tensor picked along the particle
4D-world-tube.  We devoted particular attention to the properties of
the four-dimensional counterpart of these solutions in their
Schwarzschild limit, in attempts to read the obtained results with
the experimental data. A modification of the circular stable orbits
has been investigated in agreement with the experimental
constraints. In particular, we found in both approaches that, stable
circular orbits  are possible in general in a region below the
Schwarzschild limit ($r=6M$). We therefore explored the range of
possible values of the theory parameters fixing some points in the
all range of values that should led to some possible observations.
\begin{acknowledgements}
We are grateful to the  CGM group of La Sapienza University for
helpful discussions and  precious observations during all the
work.
\end{acknowledgements}

\end{document}